\PassOptionsToPackage{table}{xcolor}
\documentclass[sigconf]{acmart}

\usepackage{hyperref}
\usepackage{amsfonts}
\usepackage{algorithmic}
\usepackage{graphicx}
\usepackage{textcomp}
\usepackage{subcaption}
\usepackage{multirow}
\usepackage{listings}
\usepackage{caption}
\DeclareCaptionType{listing}[Listing][List of Listings]

\def\BibTeX{{\rm B\kern-.05em{\sc i\kern-.025em b}\kern-.08em
    T\kern-.1667em\lower.7ex\hbox{E}\kern-.125emX}}

\copyrightyear{2026}
\acmYear{2026}
\setcopyright{cc}
\setcctype{by}
\acmConference[ICS '26]{2026 International Conference on Supercomputing}{July 06--09, 2026}{Belfast, United Kingdom}
\acmBooktitle{2026 International Conference on Supercomputing (ICS '26), July 06--09, 2026, Belfast, United Kingdom}
\acmDOI{10.1145/3797905.3807863}
\acmISBN{979-8-4007-2522-7/2026/07}

\begin{document}

\title{Harnessing MPI mutations for AI error detection}

\author{Asia Auville}
\email{asia.auville@inria.fr}
\affiliation{%
  \institution{Inria}
  \city{Bordeaux}
  \country{France}
}

\author{Tim Jammer}
\email{tim.jammer@tu-darmstadt.de}
\affiliation{%
  \institution{Technische Universität Darmstadt}
  \city{Darmstadt}
  \country{Germany}
}

\author{Eric Petit}
\email{eric.petit@intel.com}
\affiliation{%
  \institution{Intel}
  \city{Hillsboro}
  \state{Oregon}
  \country{USA}
}

\author{Pablo de Oliveira Castro}
\email{pablo.oliveira@uvsq.fr}
\affiliation{%
  \institution{University of Versailles}
  \city{Versailles}
  \country{France}
}

\author{Emmanuelle Saillard}
\email{emmanuelle.saillard@inria.fr}
\affiliation{%
  \institution{Inria}
  \city{Bordeaux}
  \country{France}
}

\author{Mihail Popov}
\email{mihail.popov@inria.fr}
\affiliation{%
  \institution{Inria}
  \city{Bordeaux}
  \country{France}
}

\begin{abstract}

MPI errors are challenging to identify despite the significant number of expert verification tools. Dynamic tools (i.e., requiring profiling) are computationally expensive and accurate in error detection, whereas static analysis (i.e., operating at source code or compilation) is computationally cheap but less accurate. Interestingly, the recent success of AI and LLMs offers an alternative to increase static analysis accuracy while preserving its low overhead. Yet current methods remain difficult to benchmark, too general,  and poorly adapted to the specific challenges of high-performance computing. 

In this paper, we investigate how AI-powered tools can efficiently and accurately detect errors in real-world MPI applications. We propose a novel MPI Mutated Dataset (MMD), constructed from MPI programs extracted from thousands of open-source GitHub projects. After sorting and filtering these files, we inject errors that realistically emulate developers' mistakes using synthetic code mutations. We leverage the dataset to train different AI models and assess their generalization capabilities against standard verification tools. We train compact multilayer perceptrons (MLP) with AST-T5 embeddings (i.e., transformer only encoder network) to detect diverse MPI errors (from the MPI-BugBench (MBB) classification) including \textit{CallOrdering} or \textit{InvalidParameter}, achieving detection accuracies up to $92.0\%$. In some cases, our method outperforms the expert prompted Qwen-3 Coder LLM and provides competitive results compared to established dynamic verification tools. We further evaluate Claude Sonnet~4.5 across different prompting strategies, showing that engineered prompts improve \textit{InvalidParam} detection accuracy from $84.8\%$ to $93.6\%$, surpassing the dynamic tool ITAC ($87.0\%$). The full agentic pipeline achieves $73.9\%$ overall multi-class accuracy and $87.6\%$ bug detection recall on MBB. This paves the way for large-scale verification efforts that can use our dataset as a foundation to thoroughly investigate each MPI error in real world applications.

\end{abstract}



\maketitle

\section{Introduction}

High Performance Computing (HPC) provides the necessary computational power to accelerate scientific discoveries and industrial innovations. To meet the high computational demand, HPC systems rely on massive parallelism. MPI is the de facto standard for large-scale distributed-memory parallelism.
Parallelizing large applications is challenging and error-prone. Developers must explicitly account for the many different ways to exchange data, including  collective, point-to-point, persistent, and one-sided communications, thus easily triggering deadlocks or race conditions when synchronizing processes.
Different approaches~\cite{osti1470989} help developers to detect and avoid those errors, including static analysis, symbolic execution, concolic testing, model checking, dynamic verification, and trace-based profiling.

Because of the large diversity of programming features and complexity of current systems, no existing tool is able to detect all possible errors. They all come with restrictions: they are focused on one programming model, they target a specific issue, they work with a specific model implementation, they require a high amount of expert involvement, or they are not yet mature (see Section~\ref{sec:mpiveriftool}).

AI is emerging as an alternative for error or vulnerability detection. The insight is to associate patterns from the source code or the compiler intermediate representations (i.e., source~\cite{code2vec}, AST~\cite{ast_t5}, or LLVM IR~\cite{ir2vec}) to existing errors~\cite{karchi2024mpierrorsdetectionusing}. LLM-based methods~\cite{bitan2025unipar} can further assist the programming experience with auto-completion or prompting. 
Yet, \textbf{there are currently no 1) AI deployed methods that can reliably detect MPI usage errors in real life applications or 2) established benchmark datasets for such evaluations}. Indeed, current models have been trained on non-specialized code-bases, thus lacking specific MPI bugs, or have only been evaluated over small systematic datasets~\cite{karchi2024mpierrorsdetectionusing, Althiban2024}. This paper addresses these two gaps: 
First, to address the lack of representative MPI bugs in the wild, we propose a method that is able to synthetically inject realistic bugs in existing MPI programs. We use it to generate a large-scale representative MPI dataset that can be leveraged by HPC and verification practitioners to evaluate different tools or AI methods. Second, we evaluate on the generated dataset different AI models to scale the detection of MPI usage errors in real world applications. 
For each code, we extract a set of features as code embedding (AST-T5~\cite{ast_t5} or IR2VEC~\cite{ir2vec}) and associate it with the corresponding error label from the dataset to train supervised classifiers. 

We assessed the resulting embedding based models generalization capabilities over codes with different origins and demonstrated how they push the boundaries of detecting MPI errors at lower cost: we achieve up to $91.5\%$ average accuracy across all types of MPI errors (based on the state of the art MBB classification~\cite{MBB}) when considering mutated validation splits. We further stressed our models by both generalizing across datasets and comparing them against dynamic state of the art verification tools (e.g., ITAC~\cite{2021:intel:itac} and MUST~\cite{hilbrich2010must}): Our MLP achieved comparable results on detecting the  \textit{InvalidParam} error. Finally, we evaluated the MPI error detection capabilities of recent LLMs (e.g., Qwen-3~\cite{Qwen3-Coder-Next}, Claude Sonnet~4.5) across different prompt engineering scenarios, thus demonstrating the benefit of both MPI focused approaches and our dataset.
The main contributions of this paper are:
\begin{itemize}
\item A code mutation method that enables practitioners to implement realistic custom MPI errors (see Section~\ref{sec:mutation-tool}).
\item A comprehensive dataset containing representative MPI errors along with the corresponding correct versions, thus paving the way for both more advanced verification tools testing and AI modeling (see Section~\ref{sec:mutated_data}).
\item A supervised Multilayer Perceptron (MLP) classifier based on AST-T5 for error detection in real applications (see Section~\ref{sec:bugdetectionmodel}). 
\item Thorough detection accuracy analysis of MPI errors across datasets with existing AI methods (see Sections~\ref{sec:limitai}), our MLP based strategy (see Sections~\ref{sec:perfs}, \ref{sec:depthstudy} and \ref{sec:MCC_manual}), and standard verification tools (see Section~\ref{sec:compare_sota}). 
\item Evaluation of different prompt strategies, including an agentic pipeline, across different datasets, illustrating the benefits of MPI focused solutions (see Sections~\ref{sec:llmexp} and \ref{sec:energy}).
\end{itemize}

\textbf{\textit{Our key idea is to create a realistic dataset for practitioners to evaluate existing verification methods, and leverage it with either the enhanced AST-T5 modeling capabilities (an embedding never used before for MPI verification) or LLMs using expert crafted prompts to produce a more generalizable AI detection model}}. Section~\ref{sec:dataset} describes the dataset, Section~\ref{sec:bugdetectionmodel} presents how we leverage it with AST-T5 embeddings or prompts to produce models that are eventually evaluated in Section~\ref{sec:results}. We discuss limitations and future directions in Section~\ref{sec:fw}.

\section{Related Works}

Our work focuses on MPI error detection, AI error detection, MPI datasets, and code mutation.

\textbf{MPI Verification Strategies.}
\label{sec:mpiveriftool}
While verification tools assist developers, they are limited in scope or detection consistency.
Static characterization offers scalable and efficient analysis. PARCOACH~\cite{huchant:hal-02390025, parcoach23} uses an LLVM pass to detect misuse of a subset of MPI communications. It avoids executing programs but reports many false positives. MC SimGrid~\cite{simgrid} is a model checker that verifies safety properties and communication determinism on top of the SimGrid framework. Yet it does not capture memory consistency errors. Such approaches remain limited due to their limited compile-time information (e.g., which send matches which receive) and the non-determinism of MPI executions. Conversely, dynamic tools such as MUST~\cite{hilbrich2010must} and  ITAC~\cite{2021:intel:itac} have a larger detection scope. Yet, they scale poorly as they require time-consuming full MPI executions: MUST uses timeouts to report some errors while ITAC generates and analyzes trace files. They also suffer from non-determinism as they depend on a specific execution order and set of input parameters.
Verification tools balance the trade-off between detection accuracy and evaluation overhead. 

\textit{Our approach raises accuracy without run overhead and minimizes the need for expert knowledge in the process of detecting MPI errors.}

\textbf{AI Verification for Parallel Programs.} Preliminary efforts showcase the potential of AI for anomaly detection in HPC~\cite{lu2026survey}, and particularly for parallel error detection. Althiban et al.~\cite{10962152} identify deadlocks and race conditions with $97\%$ accuracy using DDNs over hybrid MPI and OpenMP C++ programs. Their dataset, expanded through data augmentation using GPT-3.5~\cite{Althiban2024}, contains approximately $1,500$ code samples. The generalization of these results in real-world applications is still unclear as their validation was conducted on the same dataset as the training. In 2024, Karchi et al.~\cite{karchi2024mpierrorsdetectionusing} utilize GNN models and LLVM IR embeddings to predict MPI errors. While they conducted inter-dataset evaluations, all their experiments were limited in scope to synthetic codes, thus limiting generalization. As this is the only approach exclusively focused on MPI error detection, we compare the generalization of our respective methods in Section~\ref{IR2VEC_section}. In 2026, Valero-Lara et al.~\cite{valero2026chatmpi} introduced ChatMPI, an LLM for automatic MPI parallelization and optimization: we evaluate general LLMs with different prompt strategies in Section~\ref{sec:llmexp} and expect our mutation based dataset to offer fine-tuning opportunities for such new models. In addition, Piersall et al.~\cite{piersall2026improving} demonstrate that LLMs augmented with Few-Shot bug references and CoT reasoning significantly outperform zero-shot baselines in MPI defect detection (44\% → 77\% accuracy), generalizing across multiple LLMs including Llama2, Code Llama, and Qwen2.5-Coder. However, this approach only focuses on short and synthetic codes, and lacks an analysis on real-life MPI errors.

Beyond MPI, AI for programming and verification is an active field of research.
Pradel et al.~\cite{DBLP:journals/corr/abs-1805-11683} demonstrate that training on mutated JavaScript code generalizes to real-world bugs, indicating that learning from synthetic errors represents a viable strategy. 
Similarly, Mahyari et al.~\cite{mahyari2024harnessingpowerllmssource} show that LLMs trained on LLVM IR can detect real vulnerabilities across both synthetic and real-world datasets, even outperforming static analysis tools in certain cases. Such approaches share a common challenge: code representation. They emphasize how structured representations, such as ASTs or IR-based graphs, are critical for higher accuracy~\cite{pyda2025shortcomings}. Lee et al.~\cite{lee2025bugsweeper} further illustrate how vulnerability detection accuracy not only depends on the code representation but also on the granularity of study: only a small fraction of the code is generally involved in a vulnerability. We study the relationship between code granularity and the accuracy of our approaches in Section~\ref{sec:depthstudy}.

\textit{Our approach scales up AI for MPI error detection on larger datasets with more realistic code using mutations, for real-world applicability. We pioneer the usage of AST-T5 for MPI verification and assessed LLMs' performance by refining their scope using prompt engineering.}

\textbf{MPI Code Datasets.}
MPI-BugBench~\cite{MBB} (MBB) is the reference for MPI correctness benchmarks. It was designed to evaluate MPI verification tools by merging MBI~\cite{mbi}, MPI-CorrBench~\cite{10.1145/3431379.3460652}, and RMARaceBench~\cite{rmaracebench}. It covers all the representative MPI errors with $2,000$ synthetic codes. While their synthetic nature potentially limits the generalization of models trained on them (see Section~\ref{MBB_training}), this dataset is a prime candidate for validation as each code is either correct or contains a description of its associated error. 
Beyond verification, alternative datasets dedicated to MPI include HPCorpusMPI~\cite{schneider2024mpirigenmpicodegeneration} and  MPICodeCorpus~\cite{2023arXiv230509438S}. The former contains $16,384$ codes used to assess whether LLMs can parallelize applications with MPI while the latter is composed of around $50,000$ real world MPI programs scraped from GitHub. While numerous, these datasets do not contain information relative to the MPI correctness of the code (i.e., we assume that they are overall correct), preventing us from using them for training (see Section~\ref{datasets}).

\textit{
To overcome these limitations, we introduce the MPI Mutated Dataset, a curated dataset of GitHub-scraped MPI codes, augmented with mutations and labeled with the corresponding MPI errors.}

\textbf{Mutation Techniques for MPI Programs.}
Mutation testing is an established technique for software engineering~\cite{2011mutationsurvey,2019mutationsurvey}.
The idea is to generate modified faulty versions of a program by systematically applying small transformations that mimic common programming mistakes.
Existing mutation tools such as Mull~\cite{2018mull} are not directly applicable to our use case since they lack MPI-specific mutations.
A related technique~\cite{2020mpifaulttolerance} injects runtime faults, such as memory allocation errors that mimic full network buffers, to test the fault tolerance of an MPI implementation.
In contrast, to the best of our knowledge, no existing approach modifies source code to introduce erroneous MPI usage patterns.

\textit{To fill the gap, we propose a mutation tool that introduces diverse and realistic MPI usage errors into existing applications.}

\section{Dataset Creation}
\label{sec:dataset}

\subsection{Limitations of Existing Datasets}
\label{datasets}

MPI verification and machine learning practitioners require large and high-quality training datasets that combine: (1) a representative diversity of real-world MPI usages, (2) accurate labeling of MPI correctness or error, (3) realistic insertion of any error, and (4) preservation of the context for analyzing the patterns. No existing MPI dataset meets all these requirements.

Several curated C/C++ datasets for machine learning applications employ various labeling techniques including static/dynamic analysis and commit message filtering~\cite{lu2025icvulwelllabeledccvulnerability}. However, these datasets present two key limitations for MPI analysis: \textbf{context deficiency} and \textbf{lack of specific code}. Datasets mostly focus on functions, missing context or inherent MPI interactions that arise across processes. Moreover, the absence of MPI-specific programs in common code datasets restricts model specialization. Our preliminary filtering of The Stack~\cite{kocetkov2022stack3tbpermissively} yielded very few potentially relevant codes even before MPI-specific filtering, indicating that scaling this approach would be impractical. 

Conversely, while dedicated MPI datasets such as MPICodeCorpus and HPCorpusMPI contain numerous MPI codes, they are \textbf{challenging to  label} with errors or as correct.
Naive approaches based on static analysis or commit messages are unreliable, as commit messages often lack technical precision or are misleading~\cite{9794134}.
The complexity of parallel programming with MPI exacerbates this issue: commit messages may even misidentify the root cause of an error.
For example, in \cite{wrong-commit}, a commit message states “Fix MPI deadlock: correct tag matching in boundary exchange.” While the diff indeed modifies message tags, closer inspection shows that both the original and modified tag matchings are correct. The deadlock was actually resolved by replacing multiple \texttt{MPI\_Wait} calls with a single \texttt{MPI\_Waitall}. If all processes wait for send completion without progressing the corresponding receive, a deadlock may occur depending on the MPI implementation.
Thus, the commit message incorrectly attributes the error to tag matching, illustrating the challenge of commit-message-based labeling.

Dynamic analysis is more accurate but results in significant overhead, preventing large-scale labeling. For instance, the static approach of PARCOACH labels MBB in 2 minutes against 27 and 58 minutes for the dynamic ITAC and MUST, respectively~\cite{MBB}. 

Beyond labeling, these mostly correct codes also lack the erroneous pattern diversity needed to train robust detection models: they are mostly correct as they are scraped from GitHub. MBB~\cite{MBB} is an alternative that provides labeled and diverse codes with MPI usage errors, but its synthetic nature and structural homogeneity of short alike codes prevents model generalization (see Section~\ref{MBB_training}).

\textbf{Thus, existing datasets force models to compromise between label reliability, dataset scale, or code representativity.} 

\subsection{MPI Mutation Tool}
\label{sec:mutation-tool}

To address the previous shortcomings, we propose a code mutation tool that inserts realistic errors into correct real-world applications. Controlled mutations create diverse codes that serve as error-labeled examples, resulting in a dataset fit for verification tools benchmarking and model training.

We implement $33$ mutations, each designed to introduce a specific MPI error.
Our tool targets the following most important classes of errors, according to the MPI errors categories from the state of the art MBB classification~\cite{MBB}: 

\textbf{- LocalConcurrency}: A process accesses a memory region involved in an ongoing non-blocking or one-sided communication.

\textbf{- CallOrdering}: The order of MPI calls across processes leads to a mismatch or deadlock (e.g., processes execute different sequences of collective operations).

\textbf{- MissingCall}: One or more required MPI calls are missing (e.g., a \texttt{MPI\_Isend} not followed by an \texttt{MPI\_Wait}).

\textbf{- ParameterMismatch}: Matched MPI calls use incompatible arguments (a collective where processes do not agree on a root).

\textbf{- InvalidParam}: Individual MPI function calls receive invalid parameter values, such as negative rank or invalid communicators.

Mutations introduce these errors by altering the code. They only operate when the code pattern is compatible. If a mutation is applied, we guarantee to consistently and deterministically introduce the semantic of the error, to successfully compile the resulting code without additional errors, and to preserve the other code pieces.

Our mutation tool considers 5 out of the 8 MBB categories with multiple dedicated mutations per class to increase diversity. Our partial support of MBB classes originates from errors (e.g., RMA synchronization issues) necessitating mutations that introduce them in a non-deterministic way: current mutations are deterministic thus ensuring higher quality training data with certified errors in erroneously labeled codes.

Mutations are implemented using the library LibTooling~\cite{Libtooling} from the Clang front-end. We operate on the Abstract Syntax Tree (AST) to find suitable nodes for mutations that are then applied to their respective source code locations.
This makes it easy to mutate source code that can be compiled by the front-end, within a single compilation unit. But larger code bases are also supported using a compilation database, which can be automatically generated with CMake~\cite{cmakecmd} or a standalone tool like Bear~\cite{bear}. 

\begin{listing}[H]
\begin{lstlisting}[
    language=C,
    numbers=left,
    numbersep=4pt,
    basicstyle=\scriptsize\ttfamily,
    frame=lines,
    breaklines=true,
    tabsize=1
]
//Original MPI call
MPI_Send(buf,size,MPI_INT,dest,tag,comm);

//MPI call after mutation
MPI_Send(buf,size,MPI_DOUBLE,dest,tag,comm);
\end{lstlisting}
\caption{Introducing a \textit{ParameterMismatch} error.}
\label{lst:mutation}
\end{listing}

For example, our tool introduces a \textit{ParameterMismatch} error, by altering the data type, as shown in Listing~\ref{lst:mutation}. In this case, we use AST information to select a non-matching type.
Other (global) \textit{ParameterMismatch} errors require a more complex mutation: we can split a collective operation into two calls, and introduce a mismatch between the arguments of the collective operation over all the processes.
This transformation requires a new variable to assign processes to different variants of the collective operation, while carefully managing scope to prevent symbol collisions.
Listing~\ref{lst:mutation2} illustrates a mutation that creates a \textit{CallOrdering} error by inserting a barrier, that is executed by only one process. 

The tool also provides two additional mutations (e.g., insert broadcast call before an existing collective operation) that do not introduce errors, providing a simple mechanism for dataset expansion. We further provide an interface to control mutation placement and supports multiple mutations at distinct locations. This is implemented by using different AST Matchers that can match all applicable locations for a specific mutation. For example, an \textit{InvalidRank} error is applicable to all MPI calls that have a rank argument.

The tool is available online and provides a detailed description of its fine-grained mutations\footnote{\url{https://github.com/tudasc/mpi-mutation-tool}}. 

\begin{listing}[H]
\begin{lstlisting}[
    language=C,
    numbers=left,
    numbersep=4pt,
    basicstyle=\scriptsize\ttfamily,
    frame=lines,
    breaklines=true,
    tabsize=1
]
// Original code
 case RECEIVER: {
    MPI_Recv(&buf, 1, type, SENDER, 0, MPI_COMM_WORLD, MPI_STATUS_IGNORE);
    ...
 }
 
// Code after mutation
 case RECEIVER: {
    MPI_Barrier(MPI_COMM_WORLD);
    MPI_Recv(&buf, 1, type, SENDER, 0, MPI_COMM_WORLD, MPI_STATUS_IGNORE);
    ...
 }
\end{lstlisting}
\caption{Introducing a \textit{CallOrdering} error.}
\label{lst:mutation2}
\end{listing}

\subsection{The MPI Mutated Dataset}
\label{sec:mutated_data}

We applied our mutation tool over real-world MPI scraped applications from GitHub to ensure that our dataset (accessible online\footnote{\url{https://github.com/aauville/MPI_Mutated_Dataset}}) is diverse, contains accurate labels, and preserves the error contexts.

To focus on MPI, we only considered files containing the \texttt{\#include} \texttt{ $<$mpi.h$>$} header. The resulting \textbf{MPI Mutated Dataset (MMD)} contains around $36,000$ MPI codes: approximately $2,000$ base codes mined from top-starred GitHub repositories, and about $34,000$ mutated variants (called mutants). 
We assume base codes to be correct as they originate from starred GitHub repositories. The correctness assumption for base codes was necessary; the complexity of compiling them and the lack of context around their original intended use prevents us from being able to run tools like MUST to verify them. Yet, the noise introduced in our data from such erroneous codes falsely labeled as correct seems marginal as they are rarely considered alone for target prediction as we instead focus on error patterns, and we do not observe a significant drop in detection accuracy (LLM/MLP) when we isolate the allegedly correct codes anyway (see Section \ref{sec:results}).

Each mutation introduces either a MPI error or a MPI code that is semantically correct, and the mutated code is labeled accordingly. Indeed, similar mutations are grouped into categories that reflect the unified MBB error classification. 
We introduced correct mutations to prevent error detection by identifying mutations artefacts instead of actual errors. 
To avoid complex external dependencies, we only mutated codes for which we could generate the AST with reasonable effort. We use MMD to train and validate the subsequent models. 

\begin{figure}[]
\centering
\includegraphics[width=0.75\linewidth]{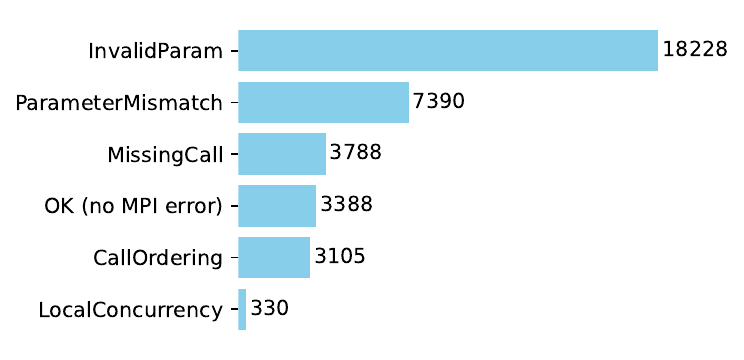}
\caption{Number of samples per MPI error.}
\label{fig:mpi_errors}
\end{figure}

Figure~\ref{fig:mpi_errors} presents MMD labels distribution. Each mutation operates over dedicated AST nodes: the number of mutated codes for a basic code is conditioned by the number of AST nodes within the code that match the mutation. Mutations corresponding to the \textit{InvalidParam} MBB error match across more AST nodes, since they are applicable at more code locations, resulting in class imbalance that we further discuss when training models in Section~\ref{sec:bugdetectionmodel}.

Table~\ref{tab:datasets} compares MPI-BugBench and MPICodeCorpus against MMD as candidates to train AI models or assess verification tools. As discussed in Section~\ref{datasets}, these datasets differ in terms of origin, labeling strategy, errors insertion, and code characteristics. MMD labels GitHub-scraped codes using the mutation process while MBB explicitly provides labels describing the MPI error for each code. On the opposite, MCC codes are not labeled. Studying MCC still remains attractive to stress our models in a different setting as it provides numerous codes assumed correct (approximation taken into account during our experiments).

Models can be directly compared on MBB or MMD benchmarks since both use the same error classification system. To guarantee fair comparison studies between MBB and MMD, we further stripped the benchmarks of any textual marker or comment outside of the code that would indicate the correctness or the nature of the code: we do not want a comment to bias the error detection process, especially in MBB. We also stripped MCC from codes found in MMD so that the two datasets do not overlap.

\begin{table}[]
\scriptsize
\centering
\caption{MPI-BugBench (MBB), MPICodeCorpus (MCC) and MPI Mutated Dataset (MMD).}
\begin{tabular}{cccccc}
\toprule
\textbf{Dataset} & \textbf{$\#$Codes} & \textbf{Origin} & \textbf{Labels} & \textbf{Real code} & \textbf{Error type}\\
\midrule
MBB & 2,000 & Synthetic & Yes & No & Generation \\
MCC & 48,000 & GitHub & No  & Yes & None\\
\textbf{MMD} & \textbf{36,000} & \textbf{GitHub} & \textbf{Yes} & \textbf{Yes} & \textbf{Mutation}\\
\bottomrule
\end{tabular}
\label{tab:datasets}
\end{table}

\section{Bug Detection Models}
\label{sec:bugdetectionmodel}

\subsection{Objective and workflow}

We aim for a model that detects genuine MPI usage errors from real-world codes.
This section presents how we designed and evaluated our different models in various scenarios with SOTA AI approaches. 

\begin{figure*}[h!]
    \centering
    \includegraphics[width=\linewidth]{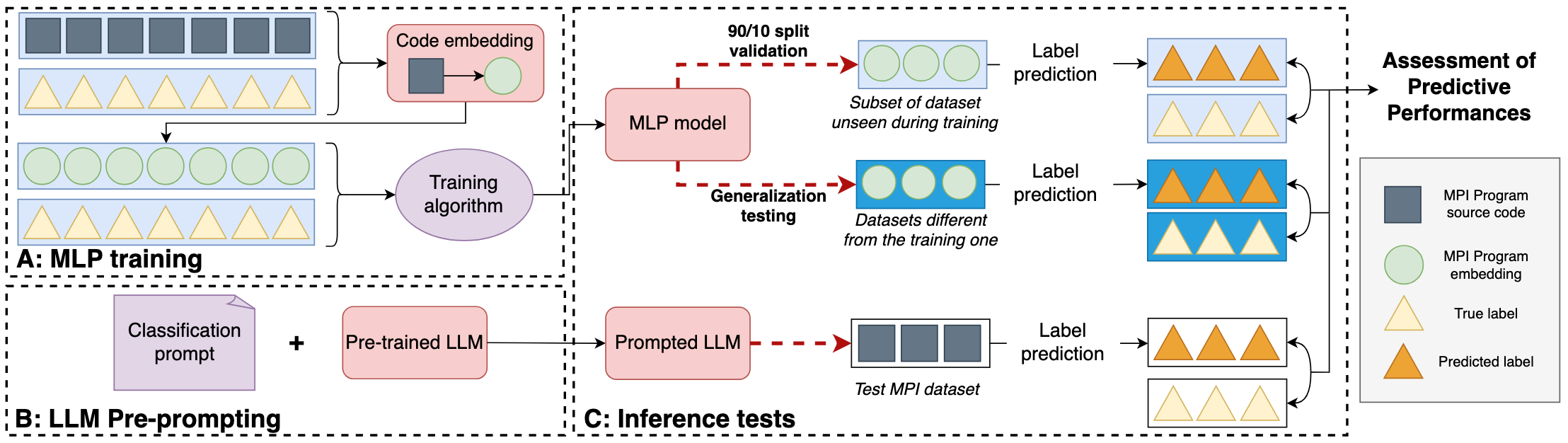}
    \caption{Global workflow of the training and testing process on the MLP model vs on the prompted LLM. Step B is further described in Figure~\ref{fig:schedprompt}.}
    \label{fig:workflow}
\end{figure*}

Figure~\ref{fig:workflow} presents the training and validation workflow we use. The process starts in \textbf{step A} by extracting for each code its respective code embedding. Pairs of embeddings and labels are then provided to an MLP model that trains to predict the error type. We consider different testing scenarios to assess the generalization capabilities of the models. Training datasets include either MMD or MBB; the former is our end model while the latter evaluates the generalization of models trained only over small synthetic codes. Beyond the datasets, we also evaluate two embedding methods (e.g., IR2VEC and AST-T5) to assess if they capture sufficient semantic and structural information to support meaningful error detection. 
\textbf{Step C} presents our inference validation test scenarios. Independently from the training dataset or the embedding, each MLP is evaluated with two strategies: split testing or over a new unseen dataset (e.g., train on MMD, validate over MBB or MCC). The former provides basic generalization tests while the latter significantly stresses the model generalization over new unseen patterns from a different origin. In parallel, \textbf{step B} illustrates how we use a large language model (LLM) prompted for error identification to compare to our MLP models. This serves to evaluate how a compact MLP compares against a general LLM model.
We further detail in Section~\ref{sec:compare_sota} how we compare these models against standard verification tools.

\subsection{Embedding Methods}

Our goal in \textbf{step A} is to represent a snippet of code with a singular vector, and use it to recognize patterns related to potential errors. For simplicity, we associate an embedding per file. We consider two embedding strategies: IR2VEC and AST-T5.

\paragraph{IR2VEC}
IR2VEC~\cite{ir2vec} operates over the LLVM Intermediate Representation (IR). Thus, IR2VEC embedding requires compiling programs through one of LLVM's front-ends, which can be tedious for some GitHub scraped programs. Indeed, many lack the context required to compile them (i.e., external libraries or dependencies). IR2VEC generates a single vector from the IR with a specific granularity, at the program, function or instruction level.
It offers a control-flow-aware analysis that we leverage at the whole program-level granularity. This embedding was considered following the promising results of \cite{karchi2024mpierrorsdetectionusing}. We will refer to models using IR2VEC embedding with the prefix \textbf{$^{IR}$}.

\paragraph{AST-T5}
AST-T5~\cite{ast_t5} is a pretraining paradigm that leverages the program's AST and generates a vectorial representation that is flow and structure aware. We use it as an encoder only, in order to generate embeddings of our programs. It was trained on open-source code and has the advantage of being usable on raw source code (i.e., no compilation is needed, unlike with IR2VEC).

\textbf{Benefits.} This model enables us to work with raw source code while benefiting from a pretraining paradigm that is AST-aware and flow-sensitive. It preserves important structural and semantic information without requiring compilation, making it compatible with a larger dataset and allowing broader coverage, improved scalability, and testing on more data while retaining the advantages of structured representations. 

\textbf{Addressing large source files.} To embed programs that exceed the tokenizer's context size, we apply a chunking method that divides the code into overlapping segments to maintain context. Specifically, we consider a chunk embedding window of $1024$ tokens that slides through the code with overlaps of $300$ tokens: this enables the model to capture context across windows. We then compute the mean of all chunk embeddings to produce a single vector for the entire program. The sliding window size for the embedding chunking mechanism was chosen as a compromise between having enough repetition to keep the context, and keeping a reasonable computing cost. We discuss the impact of this method in Section~\ref{sec:depthstudy}. In the rest of the paper, we will refer to models using AST-T5 embedding with the prefix \textbf{$^{AST}$}.

\subsection{Model Structure}
\label{sec:architecture}

This subsection describes the various models trained, during the final stage of \textbf{step A}, over distinct datasets. They all are Multilayer Perceptron (MLP) classifiers, feed-forward neural networks implemented using PyTorch's \texttt{nn.Module}. They take as input the previously described embeddings and return as output the error labels. Their architecture uses six fully connected (\texttt{Linear}) layers with progressively decreasing hidden dimensions: 512, 256, 128, 128, 64, and finally 1 for the output logit. Between the first four layers, Rectified Linear Unit (ReLU) activation functions and Dropout regularization (with a rate of 0.1) are applied to introduce non-linearity and avoid overfitting.

The structural choices were made empirically by tweaking the parameters to achieve the best performance. The goal was to keep a light-weight architecture and to study what it was able to achieve in comparison to models that are several orders of magnitude larger. 

\paragraph{\textbf{Training on MBB}}
To assess models generalization from synthetic codes and replicate the MPI verification strategy of~El Karchi et al.~\cite{karchi2024mpierrorsdetectionusing}, we trained a baseline classifier exclusively on the synthetic MBB dataset. Section~\ref{MBB_training} details the generalization capabilities of detecting \textit{InvalidParam}. We refer to these models as \textbf{$MLP_{MBB}$}.

\paragraph{\textbf{Training on MMD}}
We also trained models that we expect to have generalizable error detection capabilities over our mutated dataset. In Section~\ref{sec:perfs}, we first consider binary classification models dedicated to a specific error type (e.g., \textit{InvalidParam} or \textit{ParameterMismatch}). To further stress generalization for a unified error detection mechanism, we also trained a binary classifier between correct codes (either mutated or non-mutated) and a mixture of all possible errors introduced with mutations. Error-free mutated codes were introduced to mitigate the potential bias of the model predicting based on whether a program is a mutant or not. We further characterize the dataset labeling imbalance impact on our models in Section~\ref{sec:depthstudy} by evaluating our models with undersampling.

We refer to models trained on MMD as \textbf{$MLP_{MMD}^{correct}$} or \textbf{$MLP_{MMD}^{specific}$} (specific relating to any MBB error category including \textit{InvalidParam}, \textit{ParameterMismatch}, \textit{CallOrdering}, \textit{MissingCall}, and \textit{LocalConcurrency}). For example, \textbf{$^{AST}MLP_{MMD}^{InvalidParam}$} refers to the MLP using AST-T5, trained on MMD to recognize \textit{InvalidParam} errors.

\subsection{LLMs for MPI Error Detection}

The widespread adoption of LLMs motivates us to quantify how LLMs with different degrees of expert prompt engineering can assist in detecting MPI errors as illustrated in \textbf{Step B}.

\textbf{LLMs.} We considered two different state of the art LLMs. We evaluated Qwen-3 Coder~\cite{Qwen3-Coder-Next}, a 480B-parameter Mixture-of-Experts model with 35B active parameters, a model both tailored and considered as state of the art in $2025$ for code-related tasks. 
We further evaluated Claude Sonnet~4.5, a frontier model from Anthropic accessed via the Databricks serving API on Azure, consistently ranked among the top models for agentic coding tasks in 2025.

\subsubsection{Baseline Prompt Evaluation}
\label{sec:baseline_prompts}

We first evaluated the LLMs using independent prompts, where each prompt is applied to every code sample without orchestration. We designed two generations of prompts and compared their effectiveness.

\textbf{Initial prompts.} We manually defined a prompt for each MBB error category (\textit{Specific}), as well as a single \textit{Generic} prompt\footnote{Generic is functionally equivalent to \textbf{$MLP^{correct}$}. We still refer to it as generic as it contains light descriptions for each MBB error type.} that asks whether the code is correct or contains any error. Each Specific prompt provides examples of the target error, specifies that the model is an MPI expert, and explicitly requests reasoning behind the classification. The Generic prompt serves as a binary safe/unsafe classifier. We evaluated these initial prompts with both Qwen-3 Coder and Claude Sonnet~4.5 on MBB to establish a first baseline and enable direct model comparison on the same prompts.

\textbf{Engineered prompts.} To reduce human expertise bias in the definition of MPI error boundaries, we refined the prompting strategy using information from MBB. We first automatically generated a new prompt per error with Gemini~2.5~Pro\footnote{We used Gemini rather than Claude to avoid evaluating a model on prompts it generated.}, seeded from the original MBB publication label classification descriptions and from analysis of samples with the highest inter-class confusion. These auto-generated prompts were then manually tuned through prompt engineering to improve precision, with particular emphasis on reducing the false positive rate. Indeed, LLMs exhibit a strong tendency to over-flag errors (i.e., to ``find bugs to earn the reward''), and we counteracted this by adding explicit \textit{skepticism} instructions: the model is told to assume the code is correct unless irrefutable evidence is found. We evaluated these engineered prompts with Claude Sonnet~4.5 over MBB. Comparing the results of Claude on initial versus engineered prompts isolates the impact of prompt refinement independently from the model.

We will refer to LLM evaluations using prompt strategies as  \textbf{$^{prompt}LLM^{target}$}. For Specific prompts, the target relates to the MBB error category (e.g., \textit{InvalidParam}, \textit{CallOrdering}); for Generic prompts, the target is binary correctness. The superscript indicates the prompt generation: \textbf{I} for Initial (hand-crafted) and \textbf{E} for Engineered (Gemini-seeded/refined). For example, \textbf{$^{E}Claude^{CallOrdering}$} refers to Claude Sonnet~4.5 using the engineered prompt, aiming to detect \textit{CallOrdering} errors.

Figure~\ref{fig:schedprompt} (Initial testing) illustrates the design of the baseline prompt evaluation and their application to detect MPI errors.

\subsubsection{Agentic Pipeline}
\label{sec:agentic_pipeline}

\begin{figure*}[h!]
    \centering
    \includegraphics[width=0.86\linewidth]{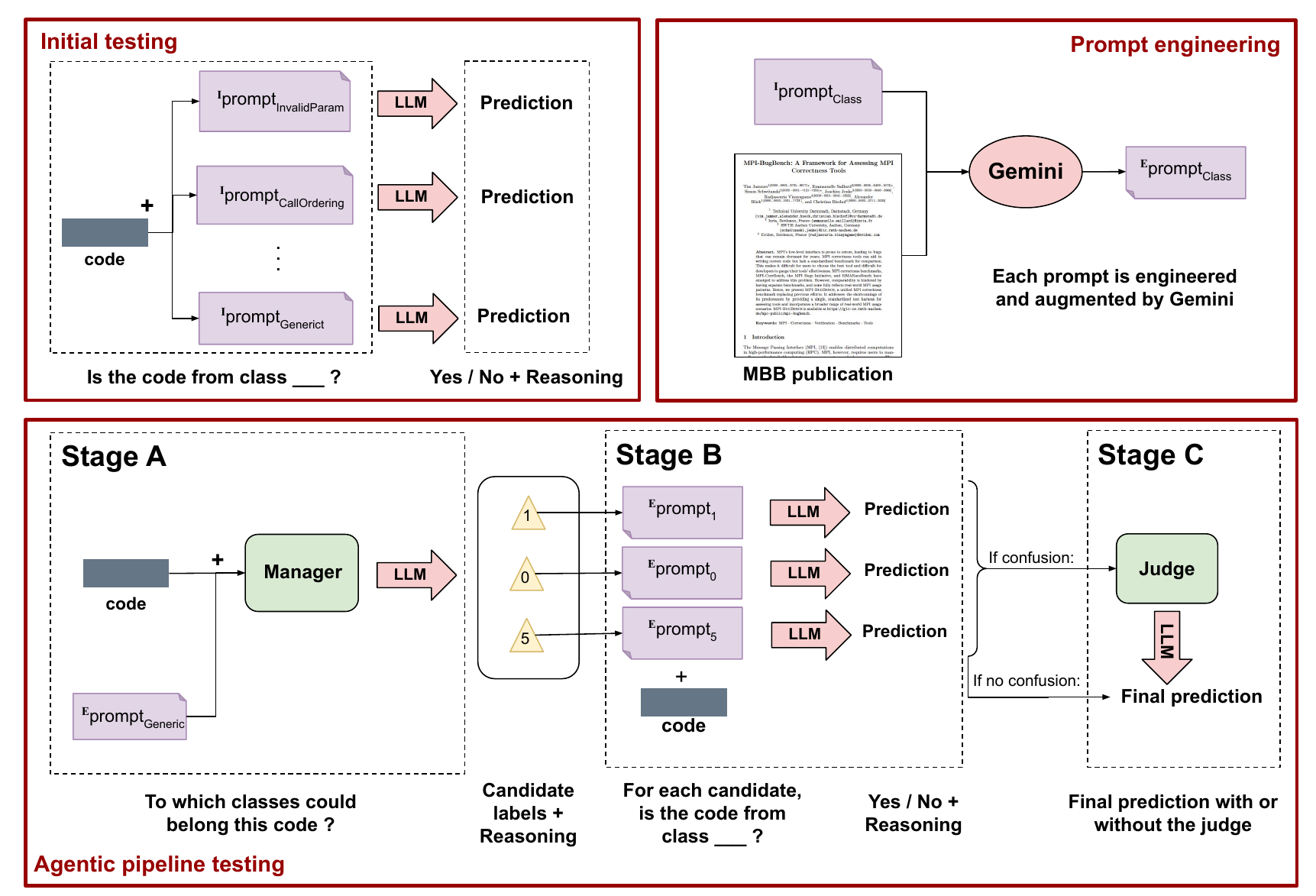}
    \caption{Initial testing: Overview of the baseline prompt evaluation strategy. Each code sample is independently evaluated by all Specific prompts (one per MBB error category) and the Generic prompt. Initial prompts ($^{I}$) provide basic error descriptions; Prompt engineering: Engineered prompts ($^{E}$) add skepticism instructions and error boundary definitions refined through inter-class confusion analysis.\\
    Agentic pipeline testing: Stage~A: the Manager screens the code and outputs a set of suspected error labels. Stage~B: only the corresponding Specialist prompts are invoked to confirm or reject. Stage~C: the Judge arbitrates when multiple Specialists confirm or the Manager expressed uncertainty.}
    \label{fig:schedprompt}
\end{figure*}

While evaluating the independent prompts, we observed that (1) the Generic prompt is accurate at detecting the presence of a bug without specific classification, and yet (2) while Specific prompts are also accurate at detecting their target error, they report many false positives, i.e., they classify other types of errors as their own. These observations motivated us to propose a multi-agent scheduling algorithm that orchestrates the prompts in a pipeline, as illustrated in Figure~\ref{fig:schedprompt} (Agentic pipeline testing).

The detection pipeline operates in three stages:
\paragraph{\textbf{Stage~A -- Manager.}} A generalist prompt first screens the code to determine whether it contains an error, and if so, identifies a group of suspected error categories. The Manager uses the Engineered Generic prompt, augmented with instructions to output a list of candidate labels. While the Manager alone is not able to accurately predict the exact error type, it reliably identifies a group that contains the actual error. Critically, the Manager can also express uncertainty by including the Safe label alongside error labels (e.g., $[0, 5]$ means ``might be Safe, but suspect CallOrdering''\footnote{$0$ and $5$ referring to correct and CallOrdering errors in the LLM output.}).

\paragraph{\textbf{Stage~B -- Specialists.}} The Manager's reasoning is forwarded only to the Specific prompts corresponding to the suspected error categories. Each Specialist either confirms or rejects the suspicion. The Specialist prompts are augmented with a \textit{skepticism} instruction: the Manager is described as ``paranoid'', and the Specialist must assume the code is Safe unless irrefutable evidence is found.

\paragraph{\textbf{Stage~C -- Judge.}} A final arbitration prompt resolves ambiguity. The Judge is invoked only when (a) multiple Specialists confirm different errors, or (b) a single Specialist confirms but the Manager expressed uncertainty (included label~0). When all Specialists reject, the code is declared Safe without invoking the Judge. When exactly one Specialist confirms and the Manager was confident, the error is accepted directly. This conditional invocation reduces API costs for clear-cut cases.

We evaluated this pipeline with Claude Sonnet~4.5 over both MBB and MMD. Results are presented in Section~\ref{sec:llmexp}.

\subsection{Model validation}
\label{sec:model_validation}
\textbf{MLP.} Figure~\ref{fig:workflow} shows the transition between \textbf{step A} and \textbf{step C}: we present the features to the models and compare the resulting predictions against the reference labels. We distinguish two validation scenarios for MLPs. We consider a standard 90\%/10\% train-validate split for models trained and evaluated over the same dataset (e.g., \textbf{$MLP_{MMD}^{correct}$} validation over MMD) to avoid overfitting. We further stress generalization by training and validating our methods over distinct \textbf{complementary datasets} (e.g., \textbf{$MLP_{MMD}^{correct}$} validation over MBB or MCC). This is a very challenging task as datasets may exhibit vastly diverse code patterns: MBB covers a large spectrum of MPI errors while MCC provides allegedly numerous correct codes.\\

\textbf{LLM.} Comparing LLMs solutions against our compact MLP highlights the latency and energy versus accuracy tradeoffs of models solely tailored for error detection against general-purpose expensive approaches. Because of the prohibitive cost of running these models, we only evaluated some scenarios. We devised the following test cases\footnote{An interesting artefact about the release of our dataset is that future production LLMs have access to it, potentially affecting their capabilities for MPI error detection.}

\textbf{(1) Baseline with initial prompts.} We evaluated both Qwen-3 Coder and Claude Sonnet~4.5 with Initial Specific and Generic prompts over MBB (\textbf{$^{I}Qwen3^{Specific}$}, \textbf{$^{I}Qwen3^{Generic}$}, \textbf{$^{I}Claude^{Specific}$}, \textbf{$^{I}Claude^{Generic}$}), enabling a direct comparison between the two models on identical prompts.

\textbf{(2) Baseline with engineered prompts.} We evaluated Claude Sonnet~4.5 with the Engineered prompts over MBB (\textbf{$^{E}Claude^{Specific}$}, \textbf{$^{E}Claude^{Generic}$} for all MBB error categories), to isolate the impact of prompt refinement.

\textbf{(3) Agentic pipeline.} We evaluated the full Manager $\rightarrow$ Specialists $\rightarrow$ Judge pipeline (Section~\ref{sec:agentic_pipeline}) with Claude Sonnet~4.5 over both MBB and MMD.\\

\textbf{Evaluation metrics.} We use standard evaluation metrics: accuracy, macro precision, macro recall, and macro F1, computed with a 0.5 decision threshold. We report macro-averaged scores because they assign equal importance to each class, which helps mitigate the impact of class imbalance. However, since class supports differ substantially in the test sets, these metrics alone do not always provide a complete picture of model performance. To better characterize the behavior of our binary classifiers on imbalanced data, we therefore also report ROC-AUC and PR-AUC scores. These metrics are threshold-independent, and PR-AUC in particular offers a more informative view of performance under class imbalance.

\textbf{Model stability.} We also applied 5 meta-repetitions for each MLP to assess prediction stability. We could not achieve this degree of robustness for the LLM evaluations due to their higher computational cost.

\section{Experimental Results}
\label{sec:results}

This section presents the prediction capabilities of the previously described models in Section~\ref{sec:bugdetectionmodel} and compares them against state of the art verification tools over the datasets from Section~\ref{sec:dataset}.

\subsection{Limits of Current AI4MPI Verification}
\label{sec:limitai} 

We compared our approaches against the state of the art method for MPI error detection using machine learning from Karchi et al.~\cite{karchi2024mpierrorsdetectionusing}. Specifically, they employ GNN or IR2VEC embedding along with decision trees to predict MPI errors trained and validated over MBB. We replicated and extended this IR2VEC based strategy to predict \textit{InvalidParam} by not only training on MBB (i.e., \textbf{$^{IR}MLP_{MBB}$}) but also on MMD (i.e., \textbf{$^{IR}MLP_{MMD}$}). 

\textbf{IR2VEC versus AST-T5.} 
\label{IR2VEC_section}
\textbf{$^{IR}MLP_{MMD}$} was instrumental in assessing the generalization capabilities of IR2VEC-based models. We used the vast code diversity from MMD for model training and then evaluated them over MBB. Table~\ref{tab:results_ir2vec} presents the prediction accuracy of this model and compares it with an AST-T5 based embedding, using \textbf{$^{AST}MLP_{MMD}$} (AST based results are further described in Section~\ref{sec:perfs}). Unlike IR based embedding, \textbf{AST-T5 demonstrates promising generalization capabilities}, hinting that it may better capture the code context needed to detect MPI errors. For these reasons, we will only refer to AST based models for the rest of the paper. AST-T5 is more recent than IR2VEC: it benefits from modern transformer architectures while being trained over order of magnitude larger datasets (i.e., the Stack~\cite{kocetkov2022stack3tbpermissively} for AST-T5 versus a few benchmark suites for IR2VEC). We hypothesize that these factors explain its superior performance.

\begin{table}[ht]
\centering
\scriptsize
\caption{Performance of \textbf{$^{IR}MLP_{MMD}^{InvalidParam}$} and \textbf{$^{AST}MLP_{MMD}^{InvalidParam}$} over MBB with 95\% confidence intervals (mean $\pm$ deviation).}
\begin{tabular}{@{}lc c@{}}
\toprule
& $^{IR}MLP_{MMD}^{InvalidParam}$ & $^{AST}MLP_{MMD}^{InvalidParam}$ \\
\midrule
\textbf{Acc}.   & 0.715 $\pm$ 0.030  & 0.791 $\pm$ 0.018 \\
\textbf{F1}     & 0.482 $\pm$ 0.024  & 0.726 $\pm$ 0.012 \\
\textbf{Prec}.  & 0.561 $\pm$ 0.069 & 0.728 $\pm$ 0.026 \\
\textbf{Rec}.   & 0.506 $\pm$ 0.022  & 0.734 $\pm$ 0.020 \\
\bottomrule
\end{tabular}
\label{tab:results_ir2vec}
\end{table}

\textbf{Generalization from Synthetic Codes.}
\label{MBB_training} 
To estimate the generalization capabilities of codes trained on synthetic codes, we evaluate \textbf{$^{AST}MLP_{MBB}$} error detection accuracy over MCC or MMD. The validation reached 100\% accuracy on MBB only after a few epochs, indicating overfitting: \textbf{$^{AST}MLP_{MBB}$} specializes in detecting the MBB synthetic classes, rather than the general patterns representing the error. This was confirmed over MMD and MCC as the model failed to outperform random predictions. This illustrates how critical MPI datasets are for AI detection: \textbf{Fully synthetic datasets fail to capture the real world application diversity}.

\subsection{Generalizing MPI Verification}
\label{sec:perfs}
\begin{table}[ht]
\scriptsize
\begin{center}
\caption{AST-T5 based MLP model accuracy on MCC with 95\% confidence intervals (mean $\pm$ deviation).}
\label{tab:results_mcc}
\begin{tabular}{cc}
\toprule
\textbf{Model} & \textbf{Accuracy} \\
\midrule
\textbf{$^{AST}MLP_{MMD}^{Correct}$} & 0.917 ± 0.012 \\
\midrule
\textbf{$^{AST}MLP_{MMD}^{InvalidParam}$} & 0.744 ± 0.017 \\
\midrule
\textbf{$^{AST}MLP_{MMD}^{ParameterMismatch}$} & 0.947 ± 0.003 \\
\midrule
\textbf{$^{AST}MLP_{MMD}^{MissingCall}$} & 0.601 ± 0.030 \\
\midrule
\textbf{$^{AST}MLP_{MMD}^{CallOrdering}$} & 0.818 ± 0.025 \\
\midrule
\textbf{$^{AST}MLP_{MMD}^{LocalConcurrency}$} & 0.928 ± 0.014 \\
\bottomrule
\end{tabular}
\end{center}
\end{table}

\begin{table*}[h]
\scriptsize
\centering
\caption{AST-T5 based MLP model performances with 95\% confidence intervals (mean $\pm$ deviation). F1, Precision, and Recall are macro-averaged scores computed at a decision threshold of 0.5. Metrics are computed on classifier-specific test sets with unequal class supports; therefore macro-F1 scores are not directly comparable across models. Threshold-independent metrics (ROC-AUC and PR-AUC) are also reported. For MMD, scores only includes samples that were unseen during training.\\ }
\begin{tabular}{cc>{\columncolor{gray!20}}cccc>{\columncolor{gray!20}}cc}
\toprule
\textbf{Model} & \textbf{Test Dataset} & \textbf{F1} & \textbf{Accuracy} & \textbf{Precision} & \textbf{Recall}  & \textbf{PR AUC} & \textbf{ROC AUC} \\
\midrule

\multirow{2}{*}{\textbf{$^{AST}MLP_{MMD}^{Correct}$}} 
& MMD & 0.863 ± 0.010 & 0.864 ± 0.009 & 0.869 ± 0.008 & 0.863 ± 0.010 & 0.938 ± 0.008 & 0.929 ± 0.009 \\
& MBB & 0.530 ± 0.037 & 0.756 ± 0.040 & 0.552 ± 0.046 & 0.540 ± 0.024 & 0.249 ± 0.030 & 0.574 ± 0.035 \\
\midrule

\multirow{2}{*}{\textbf{$^{AST}MLP_{MMD}^{InvalidParam}$}} 
& MMD & 0.915 ± 0.005 & 0.915 ± 0.004 & 0.918 ± 0.004 & 0.915 ± 0.004 & 0.982 ± 0.002 & 0.981 ± 0.002 \\
& MBB & 0.726 ± 0.012 & 0.791 ± 0.018 & 0.728 ± 0.026 & 0.734 ± 0.020 & 0.669 ± 0.016 & 0.823 ± 0.008 \\
\midrule

\multirow{2}{*}{\textbf{$^{AST}MLP_{MMD}^{ParameterMismatch}$}} 
& MMD & 0.865 ± 0.021 & 0.866 ± 0.019 & 0.877 ± 0.012 & 0.866 ± 0.019 & 0.961 ± 0.006 & 0.962 ± 0.006 \\	
& MBB & 0.407 ± 0.020 & 0.594 ± 0.007 & 0.692 ± 0.054 & 0.517 ± 0.009 & 0.544 ± 0.021 & 0.616 ± 0.014 \\			 
\midrule

\multirow{2}{*}{\textbf{$^{AST}MLP_{MMD}^{MissingCall}$}} 
& MMD & 0.869 ± 0.004 & 0.869 ± 0.004 & 0.872 ± 0.005 & 0.868 ± 0.004 & 0.939 ± 0.006 & 0.946 ± 0.004 \\
& MBB & 0.443 ± 0.020 & 0.791 ± 0.061 & 0.500 ± 0.001 & 0.496 ± 0.051 & 0.002 ± 0.0004 & 0.474 ± 0.068 \\
\midrule

\multirow{2}{*}{\textbf{$^{AST}MLP_{MMD}^{CallOrdering}$}} 
& MMD & 0.909 ± 0.012 & 0.909 ± 0.012 & 0.913 ± 0.012 & 0.909 ± 0.012 & 0.969 ± 0.007 & 0.965 ± 0.010 \\
& MBB & 0.471 ± 0.018 & 0.802 ± 0.027 & 0.471 ± 0.018 & 0.473 ± 0.020 & 0.091 ± 0.008 & 0.433 ± 0.031 \\
\midrule

\multirow{2}{*}{\textbf{$^{AST}MLP_{MMD}^{LocalConcurrency}$}} 
& MMD & 0.878 ± 0.082 & 0.880 ± 0.083 & 0.880 ± 0.071 & 0.890 ± 0.074 & 0.877 ± 0.081 & 0.931 ± 0.049 \\
& MBB & 0.521 ± 0.005 & 0.875 ± 0.033 & 0.523 ± 0.004 & 0.573 ± 0.033 & 0.077 ± 0.011 & 0.758 ± 0.041 \\
\bottomrule
\end{tabular}
\label{tab:results}
\end{table*}

Table~\ref{tab:results} presents the previously described MLP  in Section~\ref{sec:architecture} targeting the diverse MPI errors over MMD and MBB. We sampled the training data to match the less numerous labels to predict. The reported metrics are further detailed in \ref{sec:model_validation}.

\textbf{\textit{InvalidParam} prediction}. Our MLP models achieve accuracy around $80\%$ over the 2 datasets. We observe a capacity for generalization on unseen data from distinct sources, as our models were able to transfer their learning over synthetic codes or fully correct and distinct applications. 

\textbf{\textit{Correct} prediction}. While the MLP model successfully identifies correct codes in MCC (see prediction accuracy in table ~\ref{tab:results_mcc}) and  remains accurate at $86\%$ over MMD, we observe a significant F1 score drop when moving to MBB, despite accuracy staying high. Our model was not able to generalize the error detection pattern to the synthetic error description in MBB: We systematically predict the generic error label, even for correct codes. As generic errors cover a vast amount of errors (e.g., epoch life cycle, message race, global concurrency, resource leak, request life cycle, call ordering) we assume that the lack of a specific pattern significantly challenges the generalization. We tested this hypothesis by evaluating each MPI error separately and investigating if they are detected as well as \textit{InvalidParam}: we observed a similar lack of inter-dataset generalization over MBB. It is notable that for the same reasons than for MMD (see \ref{sec:mutated_data}) we considered MCC codes as all correct. This is further analyzed in section \ref{sec:MCC_manual}.

\subsection{In Depth Accuracy Characterization}
\label{sec:depthstudy}
We investigate in details the predictions of our different MLP solutions. We characterize the prediction accuracy impact from different sources including 1) amount of training data, 2) the code size, 3) the consequences of different mutations referring to the same error along with different data splitting strategies, and 4) the intrinsic difference between errors from mutations (within MMD) or from generation (with MBB). While we only present here \textit{InvalidParam} and \textit{ParameterMismatch} for space constraints, these errors are representative of accurate and inaccurate MLP prediction scenarios.

\begin{figure}
    \centering
    \includegraphics[width=1\linewidth]{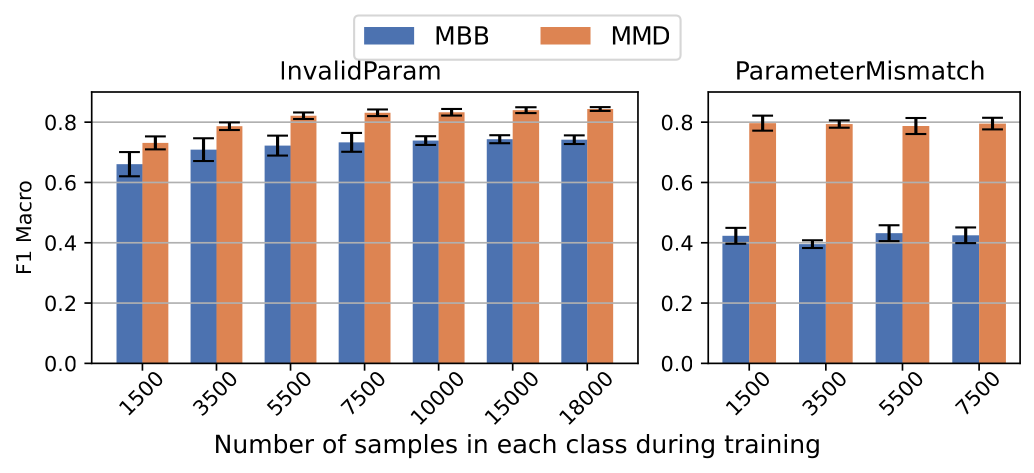}
    \caption{Impact of the amount of training data over prediction F1 score for \textit{ParameterMismatch} and \textit{InvalidParam} classifiers over the MBB and the MMD datasets.}
    \label{fig:trainamount}
\end{figure}

\textbf{Amount of training  code}. A key difference between errors such as \textit{InvalidParam} and \textit{ParameterMismatch} is the amount of codes labeled as such in MMD: \textit{InvalidParam} is the most numerous label in MMD. Thus we evaluated, by subsampling, the relationship between the amount of training codes and the detection accuracy for both errors to determine if the number of trained samples conditions generalization over MBB. Figure~\ref{fig:trainamount} demonstrates that the detection accuracy is not capped by the number of codes, indicating that the amount of training code is not a limiting factor. 

\begin{figure}
    \centering
    \includegraphics[width=0.65\linewidth]{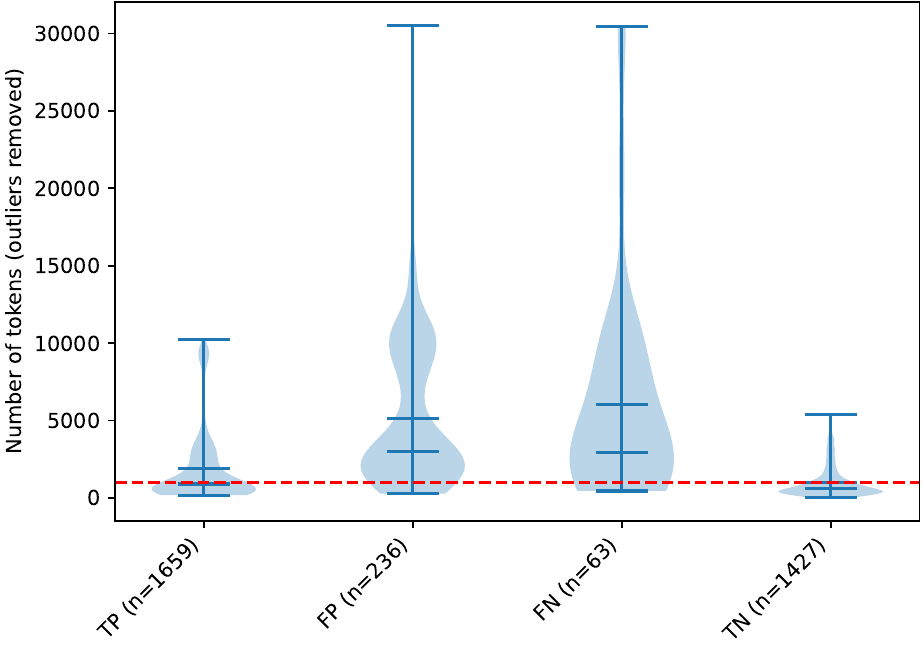}
    \caption{Impact of code size on \textit{InvalidParam} predictions. The red line shows the context size of the MLP embedding.}
    \label{fig:codesizelimit}
\end{figure}

\textbf{Code size impact on error detection.} Code length can significantly affect the detection accuracy of our models. We therefore examined the relationship between code size and model performance. Figure~\ref{fig:codesizelimit} shows the distribution density of codes across the different prediction outcomes. Most correctly predicted instances fall below the 1024-token limit, which corresponds to the embedder’s context size. This suggests that longer codes are harder to classify and appear more frequently among false positives and false negatives, possibly because the embedding process for long codes dilutes relevant information in the resulting representation vector.

\begin{figure}[h]
    \centering
    \includegraphics[width=0.8\linewidth]{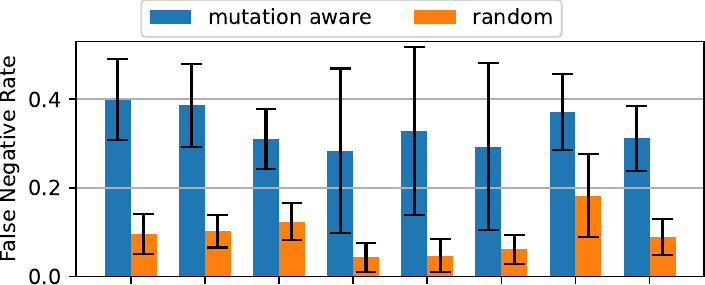}
    \caption{Rate of false negative predictions for each of the \textit{ParameterMismatch} mutations for the two validation split methods (the lower, the better).}
    \label{fig:split}
\end{figure}

\textbf{Accuracy per mutation over subsets.} 
The same MPI error can be produced by multiple mutations, thus appearing over diverse code patterns. We further measured the prediction accuracy of \textit{ParameterMismatch} per mutation on MMD to understand if some mutations are harder to identify. To stress these effects, we tried two split strategies: \texttt{mutation aware}: all original codes and their mutated variants are always kept within the same split (either training or validation), so that no code and its mutations appear across both sets; and \texttt{random}: mutated and origin codes can appear independently across both subsets. \texttt{mutation aware} mimics MBB testing with diverse codes (and shows comparable accuracy for both \textit{ParameterMismatch} and \textit{InvalidParam}) and thus guarantees that the model operates over new unseen applications while \texttt{random} offers more flexibility by potentially including an origin code in the training set and its mutated versions into the validation set: this thus forces the model to focus on the error pattern rather than the general code shape for any error prediction. For these reasons, we used \texttt{random} as our mainstream split. 
Figure~\ref{fig:split} details the false negative rate per mutation over the split strategies: \texttt{mutation aware} and \texttt{random} models uniformly predict \textit{ParameterMismatch} errors across mutations hinting that 
the lower detection accuracy of some errors is not linked to the different mutations implementing them. We also observe that
\texttt{mutation aware} split underperforms compared 
to \texttt{random aware} split suggesting the importance of the context code around the mutation.

\textbf{Error generation versus mutations.}
Inter-dataset generalization is a very challenging task. 
To evaluate the context shift between MMD and MBB, we introduced mutations into MBB and studied how classifiers trained on MMD detect those samples containing two error labels: one from MBB and one introduced by mutation.

\begin{figure}[]
    \centering
    \includegraphics[width=0.70\linewidth]{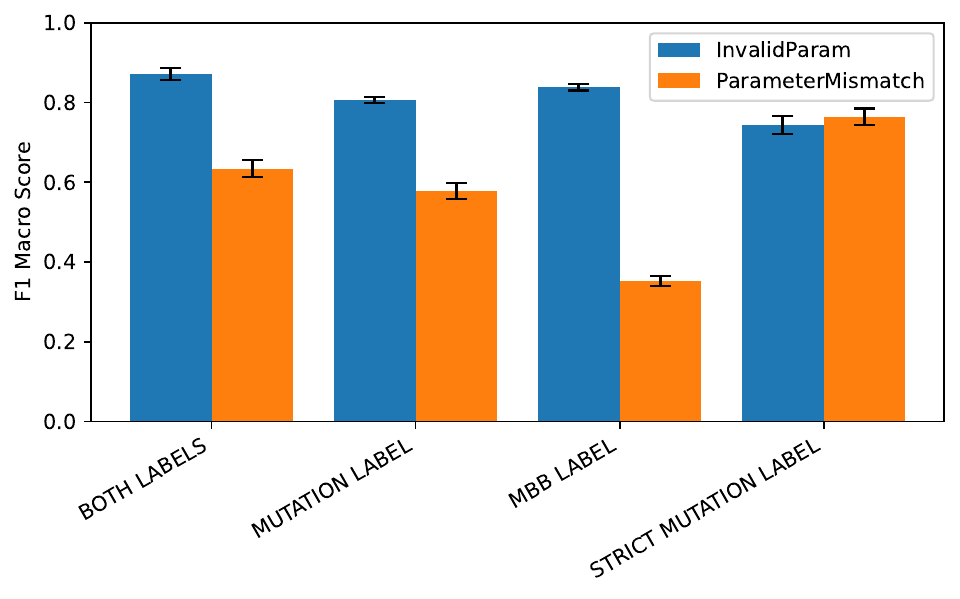}
    \caption{Impact of code mutation over MBB for \textit{ParameterMismatch} and \textit{InvalidParam} classifiers.}
    \label{fig:mbbmutation}
\end{figure}

We evaluated the classifiers applying four different paradigms to derive a single label from the two original ones (MBB and MMD). Figure \ref{fig:mbbmutation} presents the results across four different binary detection paradigms. The legend specifies which codes were considered part of the target class for each paradigm in the test set, while the other class from the binary classification consistently contained all codes that do not contain the target error, either in MBB or from mutation. \textit{Both labels} refer to codes that contain the target error in both MBB and the mutation. \textit{Mutation label} considers only the error introduced by the mutation, regardless of the original MBB label. \textit{MBB label} considers only the existing MBB error, regardless of the introduced mutation. \textit{Strict mutation label} refers to codes that were mutated with the target error starting from a correct MBB code.

The \textit{InvalidParam} classifier being the one that generalizes better on regular MBB, we observe that the scores stay fairly stable even after mutation, with the best F1 score achieved when the error is introduced twice in the code (\textit{both labels}). 
The \textit{ParameterMismatch} classifier achieves its highest performance on mutations applied to correct MBB codes (\textit{strict mutation labfel}). Accuracy drops when mutations are applied to MBB codes that already contain errors, and it performs worst when attempting to detect only the \textit{MBB label} without the help of mutations.

The performances of the \textit{InvalidParam} classifier confirm that there is no inherent flaw in the structure of the two datasets that would prevent from transferring the knowledge learned on MMD. However, the model stays confused on the detection of other errors on MBB, including \textit{ParameterMismatch}, seemingly because of the representation shift between the introduced errors.

\subsection{Manual validation of MLP predictions}
\label{sec:MCC_manual}

\begin{figure}[h]
    \centering
    \includegraphics[width=0.79\linewidth]{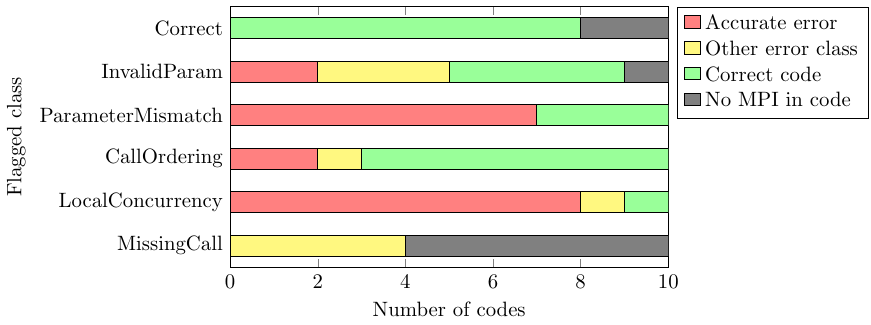}
    \caption{Actual labels of the 60 manually reviewed MCC codes depending on their MLP predicted class.}
    \label{fig:mcc_manual_test}
\end{figure}

To further evaluate our method on real-world manually reviewed codes, we tested each specialist classifier MLP trained on MMD against MCC. For each code, we obtained a prediction for each of the 6 classifiers. We then filtered the dataset to only keep codes where all classifiers agreed: one prediction is between 0.9 and 1, and all others below 0.5. This choice preselected programs that were most likely to be relevant for manual analysis. We then randomly selected 10 positive codes for each classifier, resulting in 60 codes we reviewed manually. Figure \ref{fig:mcc_manual_test} presents the actual labels of the codes flagged by each classifier. 
The 10 codes flagged as correct were indeed correct (8) or contained no MPI function (2).
Out of the 50 codes flagged as erroneous, 29 contained an MPI error, of which 19 were the exact error predicted by the specific classifier. We note that among the 60 codes, 12 were actually duplicates. In codes flagged as \textit{LocalConcurrency}, the 8 "accurate error" codes are actually 1 code and 7 exact duplicates from other repositories. 
We also observed that 9 out of 60 codes contained no MPI function, despite MCC being an MPI benchmark.

We were able to successfully leverage our small models trained on mutated code to identify real errors in an unlabeled dataset of real-life codes from Github. This also contextualizes the results in Table \ref{tab:results_mcc} regarding MCC's correctness approximation. Finally, the results emphasize the need for a labeled MPI dataset focused on correctness and data quality.

\subsection{LLMs}
\label{sec:llmexp}

We present the previously described evaluation scenarios for LLM-based MPI error detection.

\textbf{Baseline with initial prompts.}
We first evaluated both Qwen-3 Coder and Claude Sonnet~4.5 using our initial hand-crafted prompts over MBB. Table~\ref{tab:llm_baseline} presents the results for \textit{InvalidParam} and Generic classifiers. Claude outperforms Qwen-3 on both tasks: for \textit{InvalidParam}, Claude achieves higher accuracy ($84.8\%$ vs.\ $71.2\%$) driven by substantially higher recall ($0.936$ vs.\ $0.769$), though at the expense of precision ($0.629$ vs.\ $0.700$). The initial Generic prompt reveals a key challenge: Claude detects $97.7\%$ of bugs but flags $54.8\%$ of safe codes as buggy, illustrating the tendency of LLMs to over-flag errors without targeted calibration.

\textbf{Impact of prompt engineering.}
To isolate the effect of prompt refinement, we evaluated Claude Sonnet~4.5 with engineered prompts on the same MBB dataset. Results are in Table~\ref{tab:llm_baseline}.

For \textit{InvalidParam}, engineered prompts improve accuracy from $84.8\%$ to $93.6\%$ and precision from $0.629$ to $0.870$, reducing false positives by more than $4{\times}$ (FPR: $18.1\% \rightarrow 4.3\%$). Recall decreases modestly from $0.936$ to $0.870$, yielding a substantially higher F1 ($0.870$ vs.\ $0.752$). This is a clear improvement across all metrics.

For the Generic classifier, the tradeoff is more pronounced: the engineered prompt achieves near-perfect precision ($0.985$, FPR: $4.8\%$ vs.\ $54.8\%$) but at the cost of recall ($0.751$ vs.\ $0.977$), reducing overall accuracy from $87.6\%$ to $79.0\%$. The skepticism instructions make the classifier too conservative for a triage role: it misses $24.9\%$ of actual bugs, and in a pipeline with no feedback loop, a bug missed at the triage stage is permanently lost because the relevant specialist is never consulted. Conversely, the initial Generic prompt's high recall ($97.7\%$) makes it well suited as a Manager: it over-flags, but downstream specialists can filter false positives with their own skepticism. This observation directly motivated the design of our agentic pipeline (Section~\ref{sec:agentic_pipeline}), where the Manager is deliberately permissive while skepticism is delegated to the Specialists.

\begin{table}[]
\centering
\scriptsize
\caption{Baseline prompt evaluation on MBB. $^{I}$ = Initial (hand-crafted), $^{E}$ = Engineered (Gemini-seeded and refined).}
\begin{tabular}{cccccc}
\toprule
\textbf{Configuration} & \textbf{Target class} & \textbf{Accuracy} & \textbf{F1} & \textbf{Precision} & \textbf{Recall} \\
\midrule
\multirow{2}{*}{$^{I}$Qwen-3} 
 & \textit{InvalidParam} & $0.712$ & $0.688$ & $0.700$ & $0.769$ \\
 & Generic & $0.848$ & $0.648$ & $0.850$ & $0.619$ \\
\midrule
\multirow{2}{*}{$^{I}$Claude} 
 & \textit{InvalidParam} & $0.848$ & $0.752$ & $0.629$ & $0.936$ \\
 & Generic & $0.876$ & $0.927$ & $0.882$ & $0.977$ \\
\midrule
\multirow{2}{*}{$^{E}$Claude} 
 & \textit{InvalidParam} & $\mathbf{0.936}$ & $\mathbf{0.870}$ & $0.870$ & $0.870$ \\
 & Generic & $0.790$ & $0.853$ & $\mathbf{0.985}$ & $0.751$ \\
\bottomrule
\end{tabular}
\label{tab:llm_baseline}
\end{table}

\textbf{Agentic pipeline (Claude Sonnet~4.5).}
Table~\ref{tab:llm_combined_class} decomposes the per-class recall of the Claude Sonnet~4.5 agentic pipeline (Manager $\rightarrow$ Specialists $\rightarrow$ Judge, Section~\ref{sec:agentic_pipeline}) across its three stages.
Each specialist achieves high binary recall ($78$--$100\%$ Confirmed on MBB), but chaining imperfect stages compounds errors.
On MBB, this yields $73.9\%$ overall multi-class accuracy; recall ranges from $98.5\%$ (\textit{LocalParamMismatch})\footnote{We performed a fine-grained analysis of ParameterMismatch by considering LocalParamMismatch and GlobalParamMismatch separately.} to $39.7\%$ (\textit{LocalConcurrency}).

\textbf{Generalization to mutated real-world codes.}
On MMD subsampling (200~codes per class, 1400~total), the workflow achieves $65.3\%$ multi-class accuracy, with \textit{CallOrdering} at $94.5\%$ and \textit{InvalidParam} at $90.9\%$, showing that prompts designed from synthetic examples (MBB) transfer to mutation-injected real-world codes.

\begin{table}[]
\centering
\scriptsize
\caption{Per-class recall decomposition of the Claude Sonnet~4.5 agentic pipeline across MBB and MMD.
\textbf{Manager Recall}: fraction of true class-$k$ samples where label $k$ appears in the Manager's suspect list.
\textbf{Confirmed Recall}: fraction where the class-$k$ Specialist confirmed the error.
\textbf{Final Recall}: fraction where the end-to-end pipeline (after Judge) predicts class $k$. N is the number of codes in the sample.}
\begin{tabular}{clrrrr}
\toprule
\textbf{Target Class} & \textbf{Dataset} & \textbf{N}
  & \textbf{Manager} & \textbf{Confirmed} & \textbf{Final} \\
  & & & \textbf{Recall~(\%)} & \textbf{Recall~(\%)} & \textbf{Recall~(\%)} \\
\midrule
\multirow{2}{*}{\textit{Correct}}
 & MBB & 376 & --- & --- & 60.4 \\
 & MMD & 193 & --- & --- & 49.2 \\
\midrule
\multirow{2}{*}{\textit{InvalidParam}}
 & MBB & 484 & 90.3 & 80.2 & 79.3 \\
 & MMD & 197 & 95.4 & 92.4 & 90.9 \\
\midrule
\multirow{2}{*}{\textit{GlobalParamMismatch}}
 & MBB & 691 & 98.0 & 77.9 & 73.7 \\
 & MMD & 197 & 89.8 & 73.1 & 64.5 \\
\midrule
\multirow{2}{*}{\textit{LocalParamMismatch}}
 & MBB & 135 & 100.0 & 99.3 & 98.5 \\
 & MMD & 194 & 77.8 & 71.6 & 56.7 \\
\midrule
\multirow{2}{*}{\textit{MissingCall}}
 & MBB &   4 & 25.0 & 25.0 & 25.0 \\
 & MMD & 198 & 66.2 & 63.1 & 37.9 \\
\midrule
\multirow{2}{*}{\textit{CallOrdering}}
 & MBB & 215 & 86.0 & 82.8 & 81.9 \\
 & MMD & 199 & 97.5 & 96.0 & 94.5 \\
\midrule
\multirow{2}{*}{\textit{LocalConcurrency}}
 & MBB &  63 & 71.4 & 52.4 & 39.7 \\
 & MMD & 187 & 87.2 & 78.6 & 63.1 \\
\bottomrule
\end{tabular}
\label{tab:llm_combined_class}
\end{table}

\textbf{Key observations.}
Our experiments reveal five main takeaways:
\textbf{(1)}~A single initial Generic prompt (\textbf{$^{I}$Claude$^{Generic}$}) achieves high bug detection recall ($97.7\%$) but suffers from an excessive false positive rate ($54.8\%$). While unsuitable as a standalone classifier, this high recall makes it ideal as a permissive triage agent (Manager) in a multi-stage pipeline.
\textbf{(2)}~Engineered prompts dramatically reduce false positives ($4{\times}$ reduction) and for \textit{InvalidParam} specifically, \textbf{$^{E}$Claude$^{InvalidParam}$} achieves $93.6\%$ accuracy, surpassing ITAC ($87.0\%$) and MUST ($79.1\%$), demonstrating that task-specific prompt engineering can match or exceed traditional dynamic verification tools.
\textbf{(3)}~Accurately identifying the \emph{specific} error class requires the full scheduling pipeline with multiple specialist calls and Judge arbitration, as individual specialist prompts exhibit high false positive rates when applied in isolation.
\textbf{(4)}~RMA-related errors (e.g., \textit{LocalConcurrency} involving one-sided communications) remain particularly challenging for LLMs due to subtle MPI standard semantics around window synchronization.
\textbf{(5)}~The precision-recall tradeoff is role-dependent: skepticism benefits Specialists (reducing false alarms) but harms Managers (missing bugs that are then permanently lost). This highlights the need for systematic evaluation, as without MBB's diverse error categories the role-specific impact of prompt design would remain hidden.

\textbf{Importantly, the MMD dataset is what enabled this systematic comparison}: without a large, labeled, and diverse dataset of real-world MPI codes, evaluating LLM detection capabilities at scale would not have been feasible. The performance gap between MBB ($73.9\%$) and MMD ($65.3\%$) further highlights the need for realistic benchmarks when assessing AI-based verification tools.

\subsection{Comparison against verification tools}
\label{sec:compare_sota}

\begin{table}[h]
\centering
\scriptsize
\caption{Comparison of tool and LLM performances on two tasks over the MBB dataset. $^{I}$ = Initial prompts, $^{E}$ = Engineered prompts. F1, Precision, and Recall are macro-averaged scores.}
\label{tab:results_mbb_comparison}
\begin{tabular}{cccccc}
\toprule
\textbf{Tool} & \textbf{Target error} & \textbf{Accuracy} & \textbf{F1} & \textbf{Precision} & \textbf{Recall} \\
\midrule

\multirow{2}{*}{MUST} 
  & \textit{InvalidParam} & $0.791$ & $0.714$ & $0.717$ & $0.712$ \\ 
  & Generic & $0.867$ & $0.824$ & $0.793$ & $0.902$ \\ 
\midrule

\multirow{2}{*}{ITAC} 
  & \textit{InvalidParam} & $0.870$ & $0.814$ & $0.839$ & $0.796$ \\ 
  & Generic & $0.913$ & $0.876$ & $0.844$ & $0.928$ \\ 
\bottomrule

\end{tabular}

\end{table}

Table~\ref{tab:results_mbb_comparison} evaluates the dynamic state of the art verification tools MUST and ITAC over MBB. This evaluation was only conducted over MBB; evaluating dynamic verification tools on MMD is impractical. Each code requires compilation with its original dependencies and a full MPI execution, making large-scale automated evaluation prohibitively expensive (even more than LLM inference cost).

With initial prompts, Claude Sonnet~4.5 already outperforms MUST for \textit{InvalidParam} detection ($84.8\%$ vs.\ $79.1\%$), though with lower precision ($0.629$) due to a high false positive rate ($18.1\%$). Engineered prompts dramatically improve this: \textbf{$^{E}$Claude} achieves $93.6\%$ accuracy with balanced precision and recall ($0.870$), surpassing all other approaches including ITAC ($87.0\%$). For Generic error detection, ITAC retains the highest accuracy ($91.3\%$), as dynamic tools benefit from actually executing the code. However, the engineered Generic prompt achieves near-perfect precision ($0.985$) at the cost of recall ($0.751$), reflecting the skepticism-oriented design that prioritizes avoiding false alarms. \textbf{$^{AST}MLP_{MMD}^{InvalidParam}$} achieves F1 score on par with MUST ($0.73$ vs $0.71$) but remains bellow ITAC ($0.81$). These observations illustrate how our low-cost models synergize with existing expensive techniques: prompt-engineered LLMs can accurately detect specific errors and serve to guide more expensive verification mechanisms.

\begin{table}[h]
\setlength{\tabcolsep}{3pt}
\centering
\scriptsize
\caption{Methods recall comparison over MBB for all errors. LLM refers to the agentic pipeline.}
\begin{tabular}{cccccc}
\toprule
& \textit{InvalidParam} & \textit{ParameterMismatch} & \textit{MissingCall} & \textit{CallOrdering} & \textit{LocalConcurrency} \\
\midrule
MUST & \textbf{0.71} & \textbf{0.87} & 0.5 & \textbf{0.9} & 0.5 \\
\midrule
LLM & \textbf{0.79} & \textbf{0.86} & 0.25 & \textbf{0.82} & 0.4 \\
\midrule
MLP & \textbf{0.73} & 0.5 & 0.5 & 0.47 & \textbf{0.57} \\
\bottomrule
\end{tabular}
\label{tab:final_results_compare}
\end{table}

\textbf{MLP/LLM and MUST comparison over MBB} Table \ref{tab:final_results_compare} summarizes recall scores per method. AI achieves results on par with MUST, slightly outperforming \textit{InvalidParam} detection and underperforming with \textit{CallOrdering} detection, indicating synergy potential.

\subsection{Energy Measurements}
\label{sec:energy}
\textbf{Architecture.} Experiments were run on a machine with 2$\times$ NVIDIA RTX A5000 GPUs (24 GB VRAM each) and an Intel CPU with 40 physical cores (80 logical threads) and 64 GB RAM. The embedding computation was not explicitly parallelized in the code. Energy measurements were collected for both CPU and GPU.

\textbf{Results and comparison to LLM consumption.}
We analyze energy consumption across two distinct phases: training and inference. The training of our MLP required about 24.4 Wh for the largest configuration. This cost remains minimal compared to the massive energy scale required to train LLMs. As it is challenging to compute the precise cost of inference for LLMs, we rely on previous studies for rough estimations~\cite{jegham2025hungryaibenchmarkingenergy}. A single short GPT-4o query is estimated to consume 0.42Wh ($\pm$ 0.13 Wh), with 100 input and 300 output tokens. 
In contrast, the main cost in our inference comes from computing AST-T5 embeddings, measured at approximately $7.3 \times 10^{-3}$ Wh per 1000 tokens. This is around 100 times lower than a GPT-4o query.
While LLM training costs operate on a vastly different magnitude, the energy cost of inference of our proposed method is significantly lower, highlighting its efficiency for large-scale or repeated deployments. This advantage extends to latency, where the MLP pipeline requires around $0.1$s to $1$s per sample compared to around $30$s to $50$s per sample for an agentic-LLM inference.

\section{Discussions}
\label{sec:fw}

Our method takes a fundamentally distinct approach from traditional tools to detecting new errors. Traditional tools require implementing expert knowledge that describes error patterns, a process that is error-prone, requires manual labor, and is time-consuming. In contrast, our approach relies on AI: it simply needs examples to learn how to identify new errors. This shifts the costly requirement from implementing expert knowledge to labeling or generating training data. We believe that automatic mutations creating such examples have long-term potential to further reduce this cost as LLM coding capabilities improve.

A limitation to our approach is non determinism: our model returns a ratio of certainty that an error occurs. This is further emphasized by LLMs: they produce text that describes the predicted error. The format or the result itself depends on the prompt and may not match the described error topology.

Interestingly, the predicted errors can serve as a basis to elaborate a more objective error definition.
Latent text representation serves to understand the distance between words: by studying the error predictions, we can potentially define a structure to quantify the similarity between errors, using the amount of information needed to detect errors as guideline for errors classification.  

Beyond predicting errors, as future work we envision leveraging our datasets to fine-tune LLMs to not only predict errors but further fix the code. If the task proves too difficult, we can try a hybrid method with multi-source multi-model that relies on static and dynamic information (i.e., compilation and verification tools) provided to diverse models (e.g., MLP or LLMs).

\section{Conclusion}

This work investigates AI-based detection of MPI usage errors. Current solutions fail to generalize due to limited code embedding or dataset. We thus propose an alternative embedding combined with a diverse and representative dataset to train more generalizable models. This required designing a dedicated MPI mutation tool to create and augment the dataset: its mutation capabilities offer opportunities for MPI verification practitioners and for larger AI research as illustrated by our LLM pipeline. Our mutation tool, prompts, and dataset are available online.

\begin{acks}

Experiments presented in this paper were carried out using the Grid'5000 testbed, supported by a scientific interest group hosted by Inria and including CNRS, RENATER and several Universities as well as other organizations (see https://www.grid5000.fr). 
We also thank the anonymous reviewers for their feedback.

\end{acks}

\bibliographystyle{ACM-Reference-Format}

\bibliography{biblio} 

\end{document}